\documentclass[a4paper,preprint,aps,prl,superscriptaddress]{revtex4-1}
\usepackage{graphicx}
\usepackage{bm}
\usepackage{amsmath}
\usepackage{amssymb}
\usepackage{amsfonts}
\usepackage[dvips]{color}

\newcommand{\dsl}{\partial\kern-0.55em\raise 0.14ex\hbox{/}}

\newcommand{\bfk}{\bm{k}}

\begin{document}

\preprint{KYUSHU-HET-138}

\title{Convergent perturbative nuclear effective field theory}

\author{Koji Harada}
\email{harada@phys.kyushu-u.ac.jp}
\affiliation{Faculty of Arts and Science and
Department of Physics, Kyushu University\\
Fukuoka 810-8581 Japan}
\author{Hirofumi Kubo}
\email{kubo@higgs.phys.kyushu-u.ac.jp}
\affiliation{Synchrotron Light Application Center, Saga University \\
1 Honjo, Saga 840-8502, Japan}
\author{Tatsuya Sakaeda}
\email{sakaeda@higgs.phys.kyushu-u.ac.jp}
\affiliation{Department of Physics, Kyushu University\\
Fukuoka 810-8581 Japan}
\author{Yuki Yamamoto}
\email{yamamoto@koeki-u.ac.jp}
\affiliation{
Department of Community Service and Science,
Tohoku University of Community Services and Science\\
Iimoriyama 3-5-1, Sakata 998-8580 Japan}

\date{\today}

\begin{abstract}
 We consider the nuclear effective field theory including pions in the
 two-nucleon sector in the S waves up to including the
 next-to-next-to-leading order (NNLO) terms according to the power
 counting suggested by the Wilsonian renormalization group analysis done
 in a previous paper. We treat only the leading contact interaction
 nonperturbatively, and the rest, including the long-distance part of
 pion exchange, are treated as perturbations. To define the
 long-distance part, it is important to introduce a separation scale, or
 a cutoff. We employ a hybrid regularization, in which the loops with
 only contact interactions are regularized with Power Divergence
 Subtraction (PDS), while the loops with (long-distance part of) pion
 exchange are regularized with a Gaussian damping factor (GDF), to
 simplify the (nonperturbative) leading-order amplitudes. The scale
 introduced by PDS is identified with the cutoff of GDF up to a
 numerical factor. We emphasize that the introduction of the GDF
 requires a careful definition of the coupling constant for the pion
 exchange. We obtain the analytic expressions for the phase shifts for
 the $^1S_0$ and $^3S_1$-$^3D_1$ channels. By fitting them to the
 Nijmegen partial wave analysis data, it is shown that the effective
 theory expansion with perturbative long-distance part of pion exchange
 is converging.
\end{abstract}

\maketitle

\newcommand{\comment}[1]{{\color{red}\tiny #1}}

\section{Introduction}


Since the seminal papers by
Weinberg~\cite{Weinberg:1990rz,Weinberg:1991um}, nuclear effective field
theory (NEFT), an effective field theory describing systems with more
than one nucleons at low energies, has attracted much attention. See
Ref.~\cite{Epelbaum:2008ga} for a recent review. With a vast number of
papers written over twenty years, one might think that it has become a
matured discipline. In fact, there are N$^3$LO
calculations~\cite{Epelbaum:2004fk,Entem:2003ft} in the literature.


However, a very fundamental issue is still under discussions: the power
counting and renormalization. Since an effective field theory contains
an infinitely many operators, a power counting rule is necessary to
organize calculations to achieve a certain order of accuracy. Cutoff
dependence of physical quantities must be absorbed in the coupling
constants order by order, that is, consistently with the power counting.
The original Weinberg's power counting, which is widely used in most of
numerical calculations, is known to be inconsistent, i.e., cutoff
dependence arising at a certain order can be absorbed only by terms of
higher orders~\cite{Kaplan:1996nv}. Several authors
(e.g.~\cite{Nogga:2005hy}) consider variants of the Weinberg's scheme
and discuss the nonperturbative renormalization of them.


An alternative, consistent power counting scheme (known as ``KSW power
counting'') was proposed by Kaplan, Savage and
Wise~\cite{Kaplan:1998tg,Kaplan:1998we}, and independently by van
Kolck~\cite{vanKolck:1998bw}. In their scheme, pion exchanges are
treated as perturbation. Fleming, Mehen, and
Stewart~\cite{Fleming:1999ee} however showed that the effective theory
expansion fails to converge at NNLO, due to the strong tensor force of
pion exchange at short distances. Beane, Bedaque, Savage, and van
Kolck~\cite{Beane:2001bc} proposed a remedy, in which the $1/r^3$ part
of pion exchange is treated nonperturbatively.


In a series of
papers~\cite{Harada:2005tw,Harada:2006cw,Harada:2007ua,Harada:2010ba},
we advocate that the power counting should be based on the {\em scaling
dimensions} obtained by Wilsonian renormalization group (RG)
analysis. (See Refs.~\cite{Birse:1998dk,Birse:2005um,Nakamura:2004ek}
for related works.) The idea is that power counting is an order of
magnitude estimate based on the dimensional analysis, and that the
quantum notion of the dimension of an operator is the scaling
dimension. A nonperturbative RG analysis is necessary because in the S
waves the physical two-nucleon system is governed by a nontrivial fixed
point which is inaccessible in perturbation theory.


In a previous paper~\cite{Harada:2010ba}, we perform the Wilsonian RG
analysis for the nucleon-nucleon scattering in the S waves in the NEFT
including pions. We emphasize that it is important to divide pion
exchange into its short-distance part (S-OPE) and the long-distance part
(L-OPE) separated by a cutoff scale, because they play different
roles. The S-OPE is represented as contact terms. A part of the S-OPE is
relevant in the RG sense and thus should be treated nonperturbatively
while the L-OPE is treated as perturbation. It turns out that the power
counting for the contact interactions and the L-OPE is very similar to
the KSW power counting.


In this paper, we consider the NEFT including pions in the two-nucleon
sector in the S waves with the power counting suggested by the Wilsonian
RG analysis mentioned above. In order to separate the pion exchange into
two parts, we introduce an explicit separation scale, or a cutoff,
$\lambda$, and propose a new hybrid regularization in which the diagrams
with only contact interactions are regularized with Power Divergence
Subtraction (PDS)\cite{Kaplan:1998tg,Kaplan:1998we}, but those with
potential pion exchange (L-OPE) are regularized with a Gaussian damping
factor (GDF). We obtain the analytic expressions for the phase shifts
for $^1S_0$ and $^3S_1$ at NNLO and fit them to Nijmegen partial wave
analysis (PWA) data to determine the low-energy constants (LECs). All
the technical details together with the lengthy analytic expressions
will be given elsewhere.


Our approach is similar to that by Beane, Kaplan, and
Vuorinen~\cite{Beane:2008bt} (BKV) in the respect that a separation
scale is introduced. There are however important differences: (i) We use
the same regularization both for the $^1S_0$ and $^3S_1$-$^3D_1$
channels, though BKV introduce the separation scale only for the $^3S_1$
channel. (ii) We use a GDF to regularize the pion potential, while BKV
use a Pauli-Villars type regulator, which we find insufficient to render
several diagrams convergent. (iii) We interpret the separation scale as
an analog of the floating cutoff in the Wilsonian RG analysis so that it
should not exceed the physical cutoff $\Lambda_{0}\approx$ 350 MeV above
which the effective field theory description does not hold, while BKV
consider a rather large value in the range 600 MeV $\le \lambda \le$
1000 MeV, although they consider it as a low-momentum scale of
$\mathcal{O}(Q)$. (iv) We interpret the ``renormalization scale'' $\mu$
appeared in the PDS as the separation scale too, and take $\mu \sim
\mathcal{O}(\lambda)$. BKV consider that $\mu$ is independent of
$\lambda$ and choose $\mu=m_{\pi}$. (v) In our formulation, the
separation scale $\lambda$ is smaller than or equal to the physical
cutoff $\Lambda_{0}$, but otherwise arbitrary. On the other hand, BKV
tune the value of $\lambda$ to optimize the perturbation expansion.

\section{Hybrid regularization}


Since we consider that the introduction of the separation scale is
essential, we could work only with a GDF even for the contact
interactions, but such a scheme is very complicated because the leading
order (relevant) operator becomes a linear combination of several
operators when higher order operators are included. On the other hand,
in the PDS regularization the leading operator stays the same even when
the higher order operators are included. The hybrid regularization takes
advantage of both regularizations.


Our interpretation of the ``renormalization scale'' $\mu$ introduced in
the PDS as a separation scale equivalent to $\lambda$ might look
strange. It however comes from the comparison of power-divergent loop
integrations calculated in the PDS and those with an explicit
cutoff. The power divergences of the integral are represented as
polynomials of $\mu$ in the PDS.  They are typically related as
$\mu=\lambda/\sqrt{\pi}$, which is inferred by calculating the simplest
one-loop diagram with two regularizations. Note that the interpretation
is different from the usual one for the dimensional regularization such
as $\overline{MS}$ scheme in the relativistic field theory, where only
the logarithmic divergences are explicitly treated. There the scale
$\mu$ is arbitrary, but to avoid the large logarithms, it is taken to be
the typical scale of the process in question. In the present case,
however, $\mu$ is not a low-momentum scale, and a typical momentum scale
$p$ satisfies $p < \mu\alt \Lambda_{0}$.


We introduce a GDF for the pion exchange so that they represent the
L-OPE. For the scalar part, we can explicitly extract the S-OPE from
the decomposition
\begin{equation}
 \frac{\bfk^2}{\bfk^2+m_{\pi}^2}
  = 1 - \frac{m_{\pi}^2}{\bfk^2+m_{\pi}^2},
\end{equation}
where the first term may be considered as a contact interaction and the
loops containing this term are regularized with the PDS. The second term
is the L-OPE, for which we introduce a GDF $e^{-\bfk^2/\lambda^2}$.  For
the tensor part, on the other hand, because of the tensor structure, it
is impossible to extract local operators in a similar procedure. We
assume that the effects of the S-OPE is already encoded in the
coupling constants of the contact interactions, and consider only the
L-OPE.


A diagram with pion exchange may be expanded in powers of $p/\lambda$,
where $p$ is an external momentum or the pion mass. Thus a single
diagram produces a series of contributions of different orders. For
example, a diagram which appears at NLO may also contain NNLO
contributions. This is a new feature of the hybrid regularization.


A very nontrivial point with the GDF regularization comes from the
requirement that the definition of the pion exchange coupling constant
should be independent of the cutoff. In order to satisfy the
requirement, one needs to introduce an extra factor
$e^{-m_{\pi}^2/\lambda^2}$. Including the coupling constant, the scalar
part of the pion exchange may be written as
\begin{equation}
 -i \frac{g_{A}^2}{2f^2}e^{-m_{\pi}^2/\lambda^2}
  \left[
   1_{\rm PDS}
   -\frac{m_{\pi}^2}{\bfk^2+m_{\pi}^2}e^{-\bfk^2/\lambda^2}
  \right],
\end{equation}
where $1_{\rm PDS}$ stands for a contact operator that should be treated
with PDS. One can see that this definition of the coupling constant has
several favorable features: (i) The coupling constant defined by the
residue of the pole of the Yukawa potential becomes independent of the
separation scale $\lambda$. (ii) The results of loop integrals including
pion propagators with GDFs contain the factor $e^{m_{\pi}^2/\lambda^2}$,
which is canceled by the extra factor. If the extra factor is not
included, there would be (disastrous) nonlocal contributions to higher
orders when expanded in powers of $m_{\pi}^2/\lambda^2$.



The calculated amplitudes depend on $\lambda$ (and $\mu$).  All the
dependence is actually eliminated by making the LECs
$\lambda$-dependent, i.e., there is no nonlocal $\lambda$-dependence.

\section{NNLO Fitting to the Nijmegen data}


We calculate the LO, NLO, and NNLO amplitudes for the nucleon-nucleon
scattering in the $^1S_0$ and $^3S_1$-$^3D_1$ channels analytically and
fit the phase shifts obtained from the amplitudes to the Nijmegen PWA
data. The Lagrangian is the same as that of Fleming et
al.~\cite{Fleming:1999ee}.  We include diagrams containing an $S$-$D$
mixing contact term together with the diagrams given in
Ref.~\cite{Fleming:1999ee}, and calculate with the hybrid regularization
explained above. The so-called ``radiation pion'' and ``soft pion''
contributions, and the isospin breaking terms are not included. As input
parameters, we set the nucleon mass $M=938$ MeV, the pion mass
$m_{\pi}=138$ MeV, and the pion exchange coupling constant
$g^2_{A}/2f^2=4.68\times 10^{-5}$ MeV$^{-2}$. We choose
$\lambda=\sqrt{\pi}\mu=350$ MeV, which is about the physical cutoff. We
have however checked that the quality of the fitting does not depend on
the value of $\lambda$, though the fitted values of the LECs are
different for different values of $\lambda$. A detailed account based on
RG analysis will be given in a separated paper.


In order to fit the calculated phase shifts to the Nijmegen PWA data, we
use the \textsl{Mathematica} command \texttt{NonlinearModelFit} with a
weight function $\sim 1/p^{4}$, emphasizing the low-momentum region. The
range of the data varies with the order: the ranges of center-of-mass
momentum $0$ -- $30$ MeV, $0$ -- $70$ MeV, and $0$ -- $250$ MeV are used
for LO, NLO, and NNLO respectively for both the $^1S_0$ and $^3S_1$
phase shifts. If one dares to try to fit to an unreasonably wider range,
the quality of the fit in the low-momentum region becomes worse. We
should not do so because NEFT must describe the lower-momentum region
better than the higher-momentum region. Given the order of the
expansion, the range in which the fitting is successful may be viewed as
the validity region.

\begin{figure}
 \includegraphics[width=0.9\linewidth,clip]{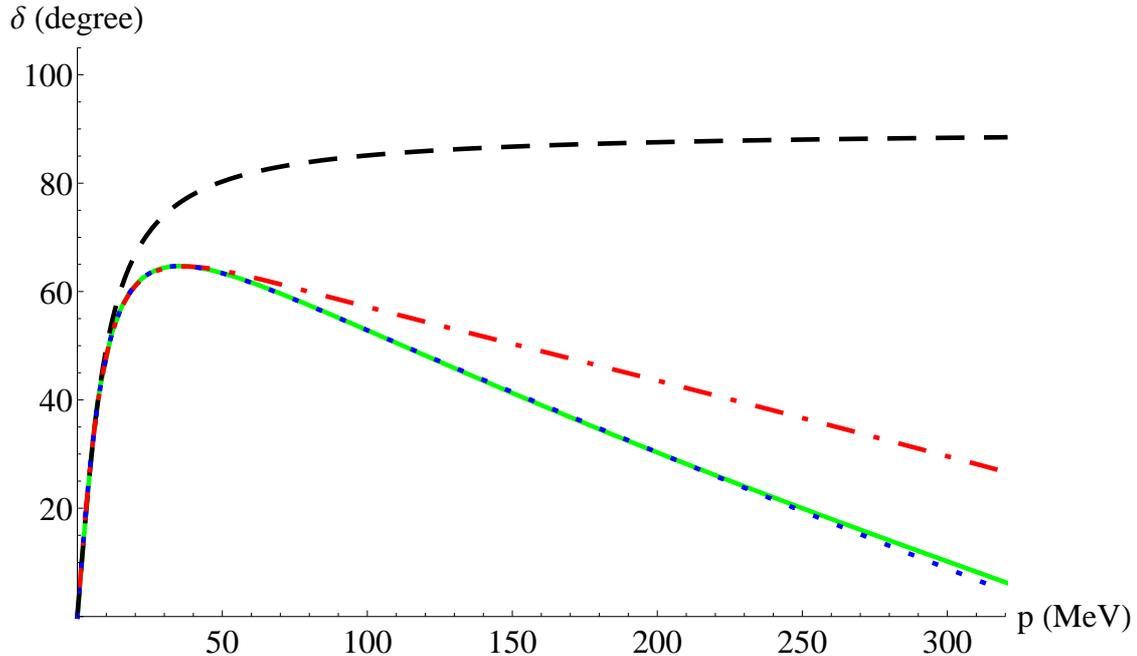}
 \caption{\label{fig:singlet} The scattering phase shift for the $^1S_0$
 channel. The LO(dashed line), NLO(dash-dotted line), and NNLO(dotted
 line) results are plotted together with the Nijmegen PWA data(solid
 line).}
\end{figure}

\begin{figure}
 \includegraphics[width=0.9\linewidth,clip]{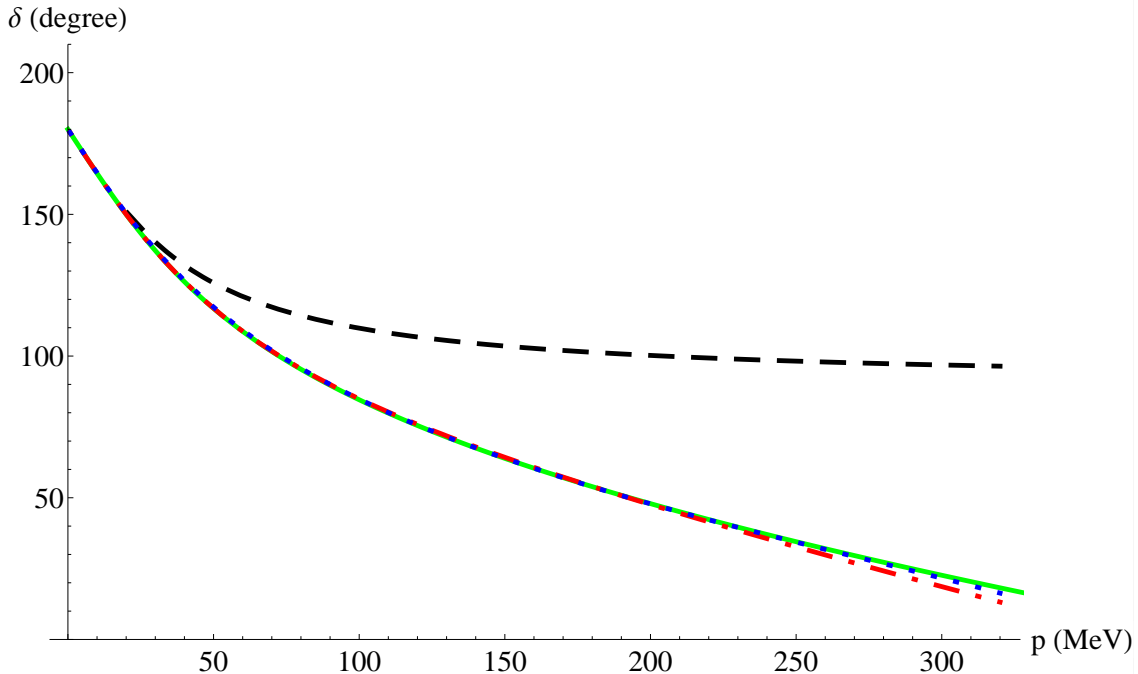}
 \caption{\label{fig:triplet} The scattering phase shift for the $^3S_1$
 channel. For the notation see Fig.~\ref{fig:singlet}.}
\end{figure}


We show the calculated scattering phase shifts by using the fitted
values of the LECs in Fig.~\ref{fig:singlet} for the $^1S_0$ channel,
and in Fig.~\ref{fig:triplet} for the $^3S_1$ channel. Most
importantly, \textit{no breakdown of the effective theory expansion is
observed.} By going to higher orders, the effective field theory fits
better and with wider validity regions.
The NNLO results are excellently fitted up to $\sim 300$ MeV.


There are one, three and six (seven) coupling constants to be fitted in
the LO, NLO and NNLO calculations respectively in the $^1S_0$
($^3S_1$-$^3D_1$) channel. We renormalize only the $\lambda$-dependence
before fitting the phase shifts to the data, in order to avoid the
cancellations among large numbers arising from the terms with positive
powers of $\lambda$.

Tables~\ref{table:singlet} and \ref{table:triplet} show how fitted
values of the ($\lambda$-renormalized) dimensionless coupling constants
change as we go higher orders.  The dimensionless coupling constants
$\tilde{X}^{(s)}_{2n}$ for $X^{(s)}_{2n}$ with $X=C, D, E$, which are
used in Ref.~\cite{Fleming:1999ee}, are defined as
$\tilde{X}^{(s)}_{2n}=(M/4\pi)\mu^{2n+1} X_{2n}^{(s)}$, where $s$ stands
for the channel.  It is important to note that most of these coupling
constants are of order one. It implies that our estimate of the
magnitude of each term is correct.
There are however large ambiguities in determining the values of
coupling constants at NNLO. There seem to be ``flat directions.''; sets
of considerably different values of the LECs give almost the same phase
shifts.

\begin{table}
\caption{\label{table:singlet}Dimensionless coupling constants fitted to
 the weighted Nijmegen PWA data in the $^1S_0$ channel.}
\begin{ruledtabular}
\begin{tabular}{c|llllll}
 & $\tilde{C}_{0}^{(^1S_0)}$ & $\tilde{C}_{2}^{(^1S_0)}$ & $\tilde{D}_{2}^{(^1S_0)}$ & 
 $\tilde{C}_{4}^{(^1S_0)}$ & $\tilde{D}_{4}^{(^1S_0)}$ & $\tilde{E}_{4}^{(^1S_0)}$ \\
 \hline 
 LO   & -0.958 & ---   & ---   & ---    & ---   & --- \\
 NLO  & -0.964 & 0.517 & 0.479 & ---    & ---   & --- \\
 NNLO & -0.967 & 0.205 & 0.099 & -0.055 & 1.180 & 1.203
\end{tabular}
\end{ruledtabular}
\end{table}

\begin{table}
\caption{\label{table:triplet}Dimensionless coupling constants fitted to
 the weighted Nijmegen PWA data in the $^3S_1$-$^3D_1$ channel.}
\begin{ruledtabular}
\begin{tabular}{c|lllllll}
 & $\tilde{C}_{0}^{(^3S_1)}$ & $\tilde{C}_{2}^{(^3S_1)}$ & $\tilde{D}_{2}^{(^3S_1)}$ & 
 $\tilde{C}_{4}^{(^3S_1)}$ & $\tilde{D}_{4}^{(^3S_1)}$ &
 $\tilde{E}_{4}^{(^3S_1)}$ &
 $\tilde{C}_{2}^{(SD)}$\\
 \hline 
 LO   & -1.22 & ---   & ---   & ---    & ---  & ---  & ---\\
 NLO  & -1.31 & 1.47  & 0.68  & ---    & ---  & ---  & ---\\
 NNLO & -1.42 & 0.89  & -0.11 & -11.72 & 4.28 & 31.1 & -4.80
\end{tabular}
\end{ruledtabular}
\end{table}



\section{Summary and Discussions}


We perform NNLO calculations for the scattering amplitudes for the
nucleon-nucleon scattering in the S waves with the power counting
suggested by the Wilsonian RG analysis done in a previous paper, which
is very similar to the KSW power counting. A novel hybrid regularization
is employed to introduce the scale $\lambda$, which separates the pion
exchange into the S-OPE and the L-OPE. We fit the calculated phase
shifts to the Nijmegen PWA data. The fitted values of most of the
coupling constants are of the natural size and the effective field
theory expansion seems converging up to including the NNLO.


In our approach, the ``renormalization scale'' $\mu$ appearing in the
PDS is identified with the scale $\lambda$ up to a numerical
constant. The scale $\lambda$ plays an analogous role to that of the
floating cutoff in the Wilsonian RG analysis. The low-momentum physical
quantities should not depend on the values of $\lambda$. This
requirement leads to a set of renormalization group equations (RGEs) for
the coupling constants. The requirement is satisfied order by
order. With the RGEs and their solutions, we have a complete control
over the $\lambda$ dependence. The RGEs and their solutions will be
given elsewhere.


We find that the introduction of the GDF requires a careful definition
of the pion exchange coupling constant and an additional factor is
necessary. This finding is possible only with analytic expressions and
the RG analysis. We think that it is also important for numerical
studies with similar damping factors.


Our approach will be applicable to other partial waves. The application
to the P waves would be of particular interest, because BKV suspect that
N$^3$LO calculations would be necessary to have a convergent
result. (Gegelia also found serious problems in the P waves in his
unpublished work.) Recently we have done a Wilsonian RG analysis for the
P waves in a similar manner to that in Ref.~\cite{Harada:2010ba}, and
have argued that the pion exchange in the P waves demotes to higher
order so that the counterterms are present to absorb the cutoff
dependence arising from the loops containing pion
exchanges\cite{Harada:2013hwa}. The calculations of the scattering phase
shifts with hybrid regularization are now in progress.

\begin{acknowledgments}
 This work is supported by JSPS KAKENHI (Grand-in-Aid for Scientific
 Research (C)) (22540286).
\end{acknowledgments}

\bibliography{NEFT}

\end{document}